\documentclass{ws-procs961x669}

\begin{document}

\title{Odd-dimensional gravitational waves from a binary system on a three-brane}

\author{D. V. Gal'tsov$^a$ and M. Khlopunov$^{a,b}$}

\address{$^a$Faculty of Physics, Lomonosov Moscow State University,\\
Moscow, 119899, Russia\\
$^b$Institute of Theoretical and Mathematical Physics, Lomonosov Moscow State University,\\
Moscow 119991, Russia\\
E-mail: galtsov@phys.msu.ru, khlopunov.mi14@physics.msu.ru}

\begin{abstract}
We consider gravitational radiation in the presence of non-compact extra dimensions. If their number is odd, all spacetime becomes odd-dimensional and formation of gravitational radiation becomes non-trivial because of violation of the Huygens principle. Gravitational waves travel with the speed of light, while the full retarded gravitational field of a localized source propagates with all velocities lower or equal to the speed of light, so special care is needed to extract radiation. Here we consider a simplified model consisting of two point masses moving on a three-brane embedded in five-dimensional bulk. Particles are assumed to interact through a massless scalar field living on the same brane, while gravitational radiation is emitted into the full five-dimensional space. We use the Rohrlich-Teitelboim approach to radiation, extracting the radiative component of the retarded gravitational field via splitting of the energy-momentum tensor. The source term consists of the local contribution from the particles and the non-local contribution from the scalar field stresses. The latter is computed using the DIRE approach to the post-Newtonian expansions. In the non-relativistic limit, we find an analog of the quadrupole formula containing the integral over the full history of motion, preceding the retarded moment of time. We compute gravitational radiation and study the orbit evolution of the non-relativistic circular binary system on the brane.
\end{abstract}


\bodymatter

\section{Introduction}

Gravitational radiation in the higher-dimensional spacetimes has become an intriguing problem in the last twenty years due to development of the gravity theories with large extra dimensions \cite{Arkani-Hamed:1998jmv,Randall:1999ee,Randall:1999vf,Dvali:2000hr}. Recent successes of the gravitational-wave astronomy give us the new tool for exploring extra dimensions \cite{Barvinsky:2003jf,Deffayet:2007kf,Andriot:2017oaz,Yu:2019jlb,Kwon:2019gsa,Corman:2020pyr}. Some constraints on their size, number and geometry have already been extracted \cite{Visinelli:2017bny,Pardo:2018ipy,Chakravarti:2019aup,Chakraborty:2017qve,Chakravarti:2018vlt,Mishra:2021waw}. Another new tool relevant to extra dimensions is shadow  of a black hole \cite{Vagnozzi:2019apd,Banerjee:2019nnj,Neves:2020doc}.

With regard to the theoretical study of multidimensional radiation, then in the majority of literary sources only radiation in even space-time dimensions is studied  \cite{Kosyakov:1999np,Cardoso:2002pa,Cardoso:2007uy,Mironov:2007nk,Kosyakov:2008wa}, while the odd dimensions 
were discussed mainly in the more academic context of the problem of radiation reaction \cite{Galtsov:2001iv,Kazinski:2002mp,Yaremko:2007zz,Shuryak:2011tt}. Huygens principle violation in odd dimensions \cite{Hadamard:book,Courant:book,Ivanenko:book} leads to   significant difference of field propagation in even and odd dimensions. In both cases, signal from the instantaneous flash reaches the observation point in time required to propagate with the speed of light. However, in odd dimensions, after the arrival of the primary signal an endless tail is observed which is not the case in the even dimensions. Mathematically, this is related to the localisation of the odd-dimensional retarded Green's functions not only on the light cone, but also inside it:
\begin{equation}
\label{eq:odd_Green_recurr}
G^{2\nu+1}_{\rm ret}(x)=\frac{(-1)^{\nu}}{(2\pi)^{\nu}}\frac{d^{\nu-1}}{(rdr)^{\nu-1}}\frac{\theta(t)\,\theta(t^2-r^2)}{\sqrt{t^2-r^2}}, \; \nu \in \mathbb{N},
\end{equation}
where $t=x^0$, and $r=|\mathbf{x}|$. The odd-dimensional gravitational radiation of the non-relativistic binary systems has been studied \cite{Cardoso:2008gn} using the effective field theory formalism \cite{Goldberger:2004jt,Goldberger:2007hy}, which is insensitive to the dimensionality of spacetime. But since it is based on the Fourier decompositions, the obtained analog of the quadrupole formula does not contain any information about the gravitational field in the wave zone and does not reveal the role of the tail in the formation of the gravitational-wave signal. So here we consider five-dimensional radiation essentially using the space-time picture.

\section{The setup}

We consider a simplified model of a binary system whose motion is confined inside the four-dimensional subspace (brane). Particles interact with each other only through the massless scalar field localised on the same brane, while gravitational radiation is five-dimensional. Our localisation mechanism is purely kinematical: if the initial conditions are restricted to the brane, and the mediator scalar field is four-dimensional, the system remains in the brane forever. Such a system admits the stable elliptical orbits, and the interaction field is free from tails, providing us with a simple setup to study the features of the odd-dimensional gravitational radiation  related to the Huygens principle violation.

Gravitational radiation of such a system can be described by the linearised theory on the Minkowski background without use of the quadratic part of the Ricci tensor corresponding to the gravitational stresses \cite{Weinberg1972}. However, one still needs to take into account the scalar field stresses corresponding to the energy of the particles interaction:
\begin{align}
\label{eq:gw_EoM}
&\square \bar{h}_{MN}(x) = - 2 \kappa_{5} \left \lbrack T_{MN}^{\rm P}(x) + T_{MN}^{\rm F}(x) \right \rbrack, \quad \partial^{M} \bar{h}_{MN} = 0, \\
\label{eq:pp_EMT}
&T_{MN}^{\rm P}(x) = \delta_{M}^{\mu} \delta_{N}^{\nu} \sum_{a=1}^{2} m_a \int d\tau_a \, \dot{z}_{a\mu} \dot{z}_{a\nu} \, \delta^{(4)} (x - z_a) \delta(x^4), \\
\label{eq:sc_EMT}
&T_{MN}^{\rm F}(x) = \frac{1}{4\pi} \delta_{M}^{\mu} \delta_{N}^{\nu} \left( \partial_{\mu} \varphi \partial_{\nu} \varphi - \frac{1}{2} \eta_{\mu\nu} \partial^{\alpha} \varphi \partial_{\alpha} \varphi \right) \delta(x^{4}),
\end{align}
where $M, N = \overline{0,4}$ and $\mu,\nu=\overline{0,3}$, and $\eta_{MN}={\rm diag} (-1,1,1,1,1)$. Here, $z_{a}^{\mu}(\tau_a)$ are the world lines of the particles and $\dot{z}_{a}^{\mu} = d z_{a}^{\mu}/d\tau_{a}$, $m_a$ are their masses, and $\kappa_5$ is the five-dimensional gravitational constant. We assume that the brane lies in the $x^4=0$ hypersurface.

The effective energy-momentum tensor of the gravitational field, by analogy with the four-dimensional theory \cite{Maggiore:book}, is given by
\begin{equation}
\label{eq:gw_EMT}
t_{MN}^{\rm rad}(x) = \frac{1}{4\kappa_{5}} \left \langle \partial_{M}\bar{h}_{ij}^{\rm TT}  \partial_{N} \bar{h}_{ij}^{\rm TT} \right \rangle, \quad i,j=\overline{1,4},
\end{equation}
where a periodic motion of the source is assumed, and bracket $\langle \ldots \rangle$ denotes  averaging over the  period. In Eq. \eqref{eq:gw_EMT}, we use the metric deviations in the transverse-traceless gauge
\begin{equation}
h_{0M}^{\rm TT} = 0, \quad h_{ii}^{\rm TT} = 0, \quad \partial^{j} h_{ij}^{\rm TT} = 0,
\end{equation}
in which   gravitational waves have five independent polarisations. Namely, polarisations of the plane wave propagating along the $x^3$-coordinate are presented by the matrix
\begin{equation}
\bar{h}_{ij}^{\rm TT} = 
\begin{pmatrix}
h_{+} - \frac{1}{2}h_{\displaystyle \circ} & h_{\times} & 0 & h_{14} \\
h_{\times} & -h_{+} - \frac{1}{2}h_{\displaystyle \circ} & 0 & h_{24} \\
0 & 0 & 0 & 0 \\
h_{14} & h_{24} & 0 & h_{\displaystyle \circ} 
\end{pmatrix},
\end{equation}
where the "cross" and "plus" polarisations are the same as in the four-dimensional theory \cite{Weinberg1972,Maggiore:book}
\begin{equation}
h_{+} = \frac{1}{2}\left( h_{11} - h_{22} \right), \quad h_{\times} = h_{12},
\end{equation}
and $h_{\displaystyle \circ}$ is the "breathing" mode \cite{Andriot:2017oaz}
\begin{equation}
h_{\displaystyle \circ} = \frac{2}{3}h_{44} - \frac{1}{3}\left( h_{11} + h_{22} \right).
\end{equation}
The brane-living observer would detect only three of them -- the standard "plus" and "cross", and the "breathing" polarisation, corresponding to the uniform shrinking and stretching of the probe masses circle lying in the plane orthogonal to the wave propagation direction \cite{Andriot:2017oaz}. We find that the breathing polarisation is non-vanishing even when both the source and an observer live on the brane.

\section{Point particles contribution}

To calculate the point particles contribution into the gravitational radiation we use the Rohrlich-Teitelboim definition of radiation \cite{Rohrlich:1961,Rohrlich:2007,Teitelboim:1970} (see also \cite{Galtsov:2004uqu,Kosyakov:1992qx,Spirin:2009zz,Galtsov:2020hhn}) based on the Lorentz-invariant decomposition of the on-shell energy-momentum tensor of the retarded field in the far zone. The Lorentz-invariant distance from the particle trajectory $z^M(\tau)$,
\begin{equation}
\hat{\rho} \equiv \hat{v}^{M} \hat{X}_{M}, \quad \hat{\rho} \xrightarrow{r \gg |\mathbf{z}|} r
\end{equation}
is equivalent to the spatial distance from the system to the observation point. Here, the following notations are used:
\begin{align}
&\hat{X}^M = x^M - z^M(\hat{\tau}) \equiv \hat{\rho} \hat{c}^M, \quad \hat{c}^M \hat{v}_M = 1, \quad \hat{c}^2 = 0, \\
&\hat{v}^M = v^M  (\hat{\tau}), \quad v^M = \frac{dz^M}{d\tau},
\end{align}
and $\hat{\tau}$ is the retarded proper time defined as
\begin{equation}
\label{eq:ret_prop_time_def}
\left(x^M - z^M(\hat{\tau})\right)^2 = \hat{X}^2 = 0, \; x^0 \geq z^0(\hat{\tau}).
\end{equation}

Then, in $D$ dimensions, the on-shell energy-momentum tensor can be expanded in the inverse powers of $\hat{\rho}$ as follows
\begin{align}
&T^{MN} = T^{MN}_{\rm Coul} + T^{MN}_{\rm mix} + T^{MN}_{\rm rad} \\
&T^{MN}_{\rm Coul} \sim \frac{A^{MN}}{\hat{\rho}^{2D-4}}, \quad T^{MN}_{\rm mix} \sim \frac{B^{MN}}{\hat{\rho}^{2D-5}} + \ldots + \frac{C^{MN}}{\hat{\rho}^{D-1}}, \quad T^{MN}_{\rm rad} \sim \frac{D^{MN}}{\hat{\rho}^{D-2}}.
\end{align}
Here, the most short-range part $T^{MN}_{\rm Coul}$ is the energy-momentum tensor of the  Coulomb part of the field; $T^{MN}_{\rm mix}$ is the mixed part, which is absent in $D=3$ and consists of more that one term in $D>4$. The remaining long-range part of the energy-momentum tensor $T^{MN}_{\rm rad}$ has the properties allowing one to associate it with the energy-momentum flux carried by the emitted part of the field. Indeed, it is separately conserved $\partial_M T^{MN}_{\rm rad} = 0$; it is proportional to the direct product of two null vectors $T^{MN}_{\rm rad} \sim \hat{c}^M \hat{c}^N$, corresponding to the propagation of the associated energy-momentum flux exactly with the speed of light $\hat{c}_M T^{MN}_{\rm rad} = 0$. Finally, it falls down as $T^{MN}_{\rm rad} \sim 1/r^{D-2}$ and gives positive definite energy-momentum flux through the distant sphere of the area $\sim r^{D-2}$.
 
These properties hold both in even and odd dimensions. However, due to the Huygens principle violation, in odd dimensions the emitted part of the energy-momentum tensor $T^{MN}_{\rm rad}$ depends on the entire history of the source's motion preceding the retarded time $\hat{\tau}$, while in even ones it is completely determined by the state of the source at this moment. Also, due to the form of the energy-momentum tensor \eqref{eq:gw_EMT}, which is the bilinear expression of the field derivatives, one can define its emitted part, by analogy with that of the energy-momentum tensor
\begin{equation}
(\partial_M \bar{h}_{ij}^{\rm TT})^{\rm rad} \sim 1/\hat{\rho}^{D/2-1}.
\end{equation}

Therefore, the radiated energy flux through the distant $(D-2)$-dimensional sphere of radius $r$ is given as
\begin{equation}
\label{eq:rad_flux}
W_{D} = \int \, T_{\rm rad}^{0 i} \; n^{i}\, r^{D-2} \, d\Omega_{D-2},
\end{equation}
where $d\Omega_{D-2}$ is the angular element, and $n^i = x^i/r$ is a unit spacelike vector in the observation direction.

Considering the gravitational radiation from the non-relativistic binary system we demonstrate that to find the non-relativistic approximation of the emitted part of the gravitational field produced by point particles one have to introduce, besides the small particles velocities $|\mathbf{v}| \ll 1$, another small parameter
\begin{equation}
\mathbf{s}(\tau) = \frac{\mathbf{z}(\hat{\tau}) - \mathbf{z}(\tau)}{\hat{\tau} - \tau}, \quad \lim_{\tau \to \hat{\tau}} \mathbf{s}(\tau) = \hat{\mathbf{v}}, \quad |\mathbf{s}| \sim |\mathbf{v}|,
\end{equation}
which is of order of particles velocities for any moment of proper time. Then, we find the point particles contribution into the gravitational radiation from the binary system as
\begin{equation}
\label{eq:pp_emit}
(\partial_M \bar{h}_{ij}^{\rm P})^{\rm rad} = - \frac{\mu \kappa_5 \bar{c}_{M}}{2^{3/2} \pi^{2} r^{3/2}} \int_{-\infty}^{\bar{\tau}} d\tau \frac{\dot{a}_i v_j + 2a_i a_j + v_i \dot{a}_j}{(\bar{\tau} - \tau)^{1/2}}, \quad \bar{c}_M = \lbrack -1, \mathbf{n} \rbrack,
\end{equation}
where $z^i = z^i_2 - z^i_1$ is the relative coordinate of the system, $\mu = m_1m_2/(m_1 + m_2)$ is its reduced mass, and $\bar{\tau} = t - r$ is the retarded proper time calculated up to the leading order. Note that the emitted part of the gravitational field \eqref{eq:pp_emit} is proportional to the null vector $\bar{c}_M$ leading to the corresponding energy-momentum tensor \eqref{eq:gw_EMT} being proportional to the direct product of two null vectors, in accordance with the Rohrlich-Teitelboim approach, and depends on the history of the system's motion preceding the retarded time $\bar{\tau}$.

\section{Scalar field contribution}

The scalar field stresses contribution into the gravitational radiation is calculated by use of the DIRE approach to the post-Newtonian expansions \cite{Pati:2000vt}. It is based on the splitting of the spacetime into the near zone, whose size is of order of the characteristic wavelength of gravitational radiation ${\cal R} = {\cal S}/|\mathbf{v}| \sim \lambda_{\rm GW}$ ($\cal S$ is the characteristic size of the binary system), and the radiation zone being exterior to the near zone.

In our calculations we assume that the observation point is in the radiation zone, the retardation of the scalar field inside the near zone is negligible, and that the energy-momentum density of the scalar field is non-vanishing only in the near zone being large enough due to the slow motion of the particles. Then, we arrive at the scalar field contribution into the gravitational radiation in the form
\begin{equation}
\label{eq:sc_emit}
(\partial_{M} \bar{h}_{ij}^{\rm F})^{\rm rad} = - \frac{\mu \kappa_{5} \bar{c}_{M}}{2^{5/2} \pi^{2} r^{3/2}} \int_{-\infty}^{\bar{\tau}} d\tau \, \frac{\ddot{a}_{i} z_{j} + 2 \dot{a}_{i} v_{j} + 2 a_{i} a_{j} + 2 v_{i} \dot{a}_{j} + z_{i} \ddot{a}_{j}}{(\bar{\tau} - \tau)^{1/2}}.
\end{equation}
As  the point particles contribution \eqref{eq:pp_emit}, this expression is proportional to the null vector $\bar{c}_M$ and depends on the history of the particles motion.

Combining the point-like   \eqref{eq:pp_emit} and the non-local scalar field \eqref{eq:sc_emit} contributions, we find the emitted part of the total gravitational field of the binary system:
\begin{equation}
\label{eq:tot_emit}
(\partial_{M} \bar{h}_{ij}^{\rm TT})^{\rm rad} = - \frac{\kappa_{5}\bar{c}_{M}}{2^{5/2}\pi^{2}r^{3/2}} \int_{-\infty}^{\bar{\tau}} d\tau \, \frac{\ddddot{Q}_{ij}^{\rm TT}}{(\bar{\tau} - \tau)^{1/2}}, \quad Q_{ij} =  \mu \left \lbrack z_{i} z_{j} - \frac{1}{4} \delta_{ij} z_k z_k \right \rbrack.
\end{equation}
As in the four-dimensional theory \cite{Maggiore:book}, it is determined by the transverse-traceless part of the quadrupole moment of the binary system but now depends on the   history of  motion.

Therefore, using the Eqs. \eqref{eq:gw_EMT} and \eqref{eq:rad_flux} we obtain the five-dimensional analog of the quadrupole formula for the angular distribution of the gravitational radiation power
\begin{equation}
\label{eq:5D_quad_form}
\frac{dW_{5}}{d\Omega_{3}} = \frac{\kappa_{5}}{128 \pi^{4}} \left \langle {\cal A}_{ij}^{\rm TT} {\cal A}_{ij}^{\rm TT} \right \rangle, \quad {\cal A}_{ij}^{\rm TT} = \int_{-\infty}^{\bar{\tau}} d\tau \, \frac{\ddddot{Q}_{ij}^{\rm TT}}{(\bar{\tau} - \tau)^{1/2}}.
\end{equation}
depending on the entire history of the binary system's motion preceding the retarded time $\bar{\tau}$. Recently, the analogous formula was obtained by use of the spectral decompositions of the retarded Green's functions \cite{Chu:2021uea}, which are insensitive to the dimensionality of the spacetime, and can be considered as a confirmation of our result.

\section{Binary system on a circular orbit}

As the simplest application of the obtained five-dimensional quadrupole formula, we consider the gravitational radiation from the non-relativistic binary system on the circular orbit. Integrating the angular distribution \eqref{eq:5D_quad_form}, we find the total gravitational radiation power of the system on the circular orbit
\begin{equation}
W_{5}^{\rm circ} = \frac{5}{9\pi} \kappa_{5} \mu^{2} R_{\rm s}^{4} \omega_{\rm s}^{7},
\end{equation}
where $R_{\rm s}$ is the orbital radius of the system, and $\omega_{\rm s}$ is its frequency of the orbital motion. Also, using the energy conservation law
\begin{equation}
\frac{dE_{\rm tot}}{dt} = - W_{5}^{\rm circ},
\end{equation}
where $E_{\rm tot}$ is the total mechanical energy of the point particles, we analyse the quasi-circular orbit shrinking of the binary system arriving at the evolution laws for the orbital frequency and radius
\begin{align}
\label{eq:freq_evol}
&\omega_{\rm s}(t) = \left( \frac{9\pi}{55\kappa_5} \right)^{3/11} \left( \sqrt{\mu} g_1 g_2 \right)^{-2/11} (t_{\rm coal} - t)^{-3/11}, \\
\label{eq:rad_evol}
&R_{\rm s}(t) = R_{0} \bigg( \frac{t_{\rm coal} - t}{t_{\rm coal} - t_{0}} \bigg)^{2/11},
\end{align}
where $t_{\rm coal}$ is the moment of time corresponding to the coalescence of the binary system $R_{\rm s} \to 0, \, \omega_{\rm s} \to \infty$, and $R_{\rm s}(t_0)=R_0$. Note that the obtained Eqs. \eqref{eq:freq_evol} and \eqref{eq:rad_evol} differ significantly from their four-dimensional analogs
\begin{equation}
\omega_{\rm s}^{\rm 4D}(t) \sim (t_{\rm coal} - t)^{-3/8}, \quad R_{\rm s}^{\rm 4D}(t) \sim (t_{\rm coal} - t)^{1/4},
\end{equation}
what could be expected, given the five-dimensional nature of gravity and infinite size of extra dimension in our model.

\section*{Acknowledgements}

The work of M. Kh. was supported by the “BASIS” Foundation Grant No. 20-2-10-8-1. The work of D.G. was supported by the Russian Foundation for Basic Research on the project 20-52-18012, and the Scientific and Educational School of Moscow State University "Fundamental and Applied Space Research".

\end{document}